# Closed Loop Superparamagnetic Tunnel Junctions for Reliable True Randomness and Generative Artificial Intelligence

Dooyong Koh[1*†], Qiuyuan Wang[1†], Brooke C. McGoldrick[1,2], Chung-Tao Chou[1,3], Luqiao Liu[1*], and Marc A. Baldo[1*]

[1]Department of Electrical Engineering and Computer Science, Massachusetts Institute of Technology, Cambridge, MA, USA.

[2]Present address: Cox Automotive, Oklahoma City, OK, USA.

[3]Department of Physics, Massachusetts Institute of Technology, Cambridge, MA, USA.

* E-mail: dooyong@mit.edu (D.K.), luqiao@mit.edu (L.L.), baldo@mit.edu (M.A.B.)

†These authors contributed equally: Dooyong Koh, Qiuyuan Wang

# Abstract

Physical devices exhibiting stochastic functions with low energy consumption and high device density have the potential to enable complex probability-based computing algorithms, accelerate machine learning tasks, and enhance hardware security. Recently, superparamagnetic tunnel junctions (sMTJs) have been widely explored for such purposes, leading to the development of sMTJ-based systems; however, the reliance on nanoscale ferromagnets limits scalability and reliability, making sMTJs sensitive to external perturbations and prone to significant device variations. Here, we present an experimental demonstration of closed loop three-terminal sMTJs as reliable and potentially scalable sources of true randomness in the field-free regime. By leveraging dual-current controllability and incorporating feedback, we stabilize the switching operation of superparamagnets and reach cryptographic-quality random bitstreams. The realization of controllable and robust true random sMTJs underpin a general hardware platform for computing schemes exploiting the stochasticity in the physical world, as demonstrated by the generative artificial intelligence example in our experiment.

# Main

Modern computational systems can derive significant benefit from hardware sources of random signals with desired statistical properties. Aside from the well-known examples of probabilistic models [1] and cybersecurity [2], the recent rapid development of new artificial intelligence (AI) algorithms [3,4] further enhances the demand for efficient hardware with stochastic functionalities [5,6]. Traditional devices and circuits based on complementary metal-oxide-semiconductor (CMOS) technology are highly successful in deterministic computing tasks but less efficient sources for random numbers. Recently, devices that exploit physical stochasticity have been developed as alternative solutions with superparamagnetic tunnel junctions (sMTJs) attracting particular interest [7-10]. Compared to other random number generator (RNG) technologies based on CMOS [11-13] or emerging-devices [14-16], sMTJs possess low energy barriers against thermal fluctuation ($E_b < 20k_BT$ with $k_B$ and $T$ being the Boltzmann constant and working temperature), superior speed, high compactness, and good energy-efficiency, while maintaining CMOS compatibility. Without requiring external triggering or seedling signals [17], the randomness generation in sMTJs places low demands on power consumption and circuit complexity.

Despite initial success in fields such as machine learning, combinatorial optimization and statistical inference [7,8,18,19], several critical challenges still exist in sMTJs, hampering their utilization in large scale circuits. Firstly, their low energy barrier and diminishing coercivity increases sensitivity to environmental parameters. Small field changes on the order of the earth's magnetic field can lead to a large drift in their switching probability. Similarly, fluctuations in the temperature can also modulate the dwell time in the two states, compromising reliability. Moreover, compared with thermally stable magnetic tunnel junction (MTJ) devices for memory applications [20], the device variations introduced by nanofabrication processes present a much more severe challenge to sMTJs, due to their low $E_b$ and small device area, resulting in different stray fields in the free magnetic layer [21] and different switching probabilities. Therefore, to obtain consistent performance, an external magnetic field must be applied and customized for individual devices, an unrealistic task for highly integrated circuits. In previous studies, mitigating algorithms and circuits including the noise whitening method using XOR gates have been proposed to pin the probability at 50% [22]. Unfortunately, besides requiring more devices, these approaches only work when the composing sMTJs operate within a narrow stochastic window, a condition that can be hard to satisfy in deeply scaled nanometer size devices.

In addition to demonstrating reliable switching probabilities, stochastic devices require further specifications in different applications. In probabilistic systems such as Ising models and Bayesian networks, RNGs must offer the ability to generate binomial distributions with certain specified probability [5]. At the same time in the domains of deep learning and generative AI, there is a substantial need for random numbers that follow a normal distribution, due to its mathematical convenience [4]. In cybersecurity, true random number generators (TRNGs) are indispensable for safeguarding the security of cryptographic systems, where even minimal predictability could undermine security [2,6,10]. A versatile, stochasticity-generation device that can be customized for different applications is therefore highly desirable.

In response to these challenges, this work presents a general solution for realizing TRNGs with sMTJs under a magnetic-field-free regime that promises to dramatically boost the scalability and reliability of sMTJ-integrated systems. In our work, we experimentally stabilize the operation of three-terminal sMTJs under various fluctuating environments with electrical feedback control and demonstrate that systems based on sMTJ TRNGs are robust for a wide range of advanced computing applications, such as cryptography and generative AI. The resulting real-time, single-sMTJ TRNG passes all tests in the NIST Statistical Test Suite and is used to successfully demonstrate generative AI. Both attributes demand controllable probability in real-time over an extended period. Our work focuses on stabilizing the operation of the sMTJs, ensuring that sMTJ-based systems are not only operational but also fundamentally secure and efficient.

**Superparamagnetic Tunnel Junctions and Dual Current Controllability**

We fabricate three-terminal in-plane sMTJs with the material stack depicted in Fig. 1(a) using electron-beam lithography and ion beam etching, with the multi-layer film consisting of (from bottom to top) Ta (14 nm)/CoFeB (1.3 nm)/MgO (1 nm)/CoFeB (2.6 nm)/Ta (5 nm)/Ru (5 nm). A range of nanopillars with diameters between 60 and 100 nm are built, with a typical scanning electron micrograph (SEM) image shown in the inset of Fig. 1(a). These sMTJs exhibit a double-well energy landscape as shown in Fig. 1(b), with the two stable states of parallel (P) and antiparallel (AP) representing digital '0' and '1'. The low energy barrier between the two states enables the device to randomly switch between two states at room temperature, as illustrated in the magnetic field dependence of bipolar stochastic fluctuations in Fig. 1(c) and the telegraph noise signal in the time trace of the MTJ resistance in Fig. 1(d).

In three-terminal sMTJs, stochastic switching properties can be modified by the two independent control knobs: the biasing current along the tantalum channel ($I_{SOT}$) and the applied current across the MTJ ($I_{MTJ}$), as shown in Fig. 1(a). According to the Néel-Arrhenius law [23], the retention time in either of the two states (see Fig. 1(d)) follows an exponential distribution, as illustrated in Fig. 1(e). The time constant for the exponential distributions, termed dwell times ($\tau_P$, $\tau_{AP}$), is governed by the equation [23-26]:

$$\tau_{P(AP)} = \tau_0 \exp\left(\frac{E_b}{k_B T}\left(1 \pm \frac{H - H_s}{H_k}\right)^2 \left(1 \mp \frac{I_{ST}}{I_{ST}^0}\right)\right), \tag{1}$$

where $\tau_0$ represents the attempt time, $H$, $H_s$ and $H_k$ are the external in-plane magnetic field, the stray field and the anisotropy field, separately. The spin torque (ST) current and the ST critical switching current are denoted as $I_{ST}$ and $I_{ST}^0$, respectively. From Equation (1), one can see that sMTJs are susceptible to external magnetic field drift when $H_k$ is low. Given the challenges of nanoscale field control compounded by device variation in the stray field $H_s$ and the anisotropy field $H_k$, tuning the external magnetic field for controlling switching probability is not a feasible solution for operating individual sMTJs reliably and uniformly in integrated sMTJ systems. Nevertheless, spin-orbit torque (SOT) [27] or spin-transfer torque (STT) currents can be used to locally tune the asymmetry in the two energy wells (see Fig. 1(b)) and change the probability of the AP state ($p_{AP} = \tau_{AP}/(\tau_P + \tau_{AP})$) [28,29]. Meanwhile, the relative ratio of $E_b/k_B T$, which directly influences the fluctuation frequency, can be varied by adjusting the barrier height or temperature, for example, through voltage-controlled magnetic anisotropy (VCMA) effect or simple current induced heating [14,26,30,31].

To exploit the dual-current controllability of digital probabilistic bits, we characterize the sMTJ's switching behavior under the influence of $I_{SOT}$ and $I_{MTJ}$. Figure 2(a) summarizes the probability of the AP state $p_{AP}$ under various $I_{SOT}$ and $I_{MTJ}$ combinations. Under fixed $I_{MTJ}$ biases, $p_{AP}$ decreases from 100 % to 0 % when $I_{SOT}$ increases in the positive direction, consistent with the SOT mechanism. Figure 2(c) shows a specific example for $I_{MTJ} = 0.024$ mA, where the sigmoid curve shape agrees with the expected behavior from $p_{AP}$'s expression and Eq. (1). Meanwhile, due to the high resistance-area product, the STT effect in our devices is largely negligible, as reflected by $p_{AP}$'s weak dependence on $I_{MTJ}$. Beside the probability, we also measure the characteristic dwell time ($\tau = \sqrt{\tau_P \tau_{AP}}$) under the two currents (see Fig. 2(b) and Fig. 2(d)). Overall, the dwell time is observed to be shorter under elevated $|I_{SOT}|$ and $|I_{MTJ}|$, mostly attributable to the Joule heating effect. The results in Fig. 2 therefore demonstrate that we possess two non-orthogonal control mechanisms: $I_{SOT}$ influences the probability through the SOT effect, and meanwhile both currents affect the fluctuation speed via Joule heating though with different levels of sensitivity. The availability of the two control knobs therefore provides the opportunity to modify the stochastic properties of sMTJ post fabrication, as well as to reach reliable and scalable RNG via feedback control, as will be discussed in the following.

**Reliable Random Number Generation**

Feedback control is a well-known solution for overcoming external perturbations and maintaining long-term stability in complex systems. Similarly, probability locked loops (PLLs) have been simulated [32,33] for thermally stable MTJ-based systems that generate and operate on stochastic signals. Implementations of PLLs, however, require controllable hardware such as our three terminal sMTJs. Figure 3(a) shows the printed circuit board (PCB) interface between the fabricated sMTJ chip and a microcontroller unit (MCU) used to operate a PLL. The controlling commands from the MCU are converted into terminal voltages $V_1$ and $V_2$ via digital-to-analog converters (DAC), which further provide the input currents of $I_{SOT}$. To read out the magnetic states, the voltage division between the sMTJ and an NMOS transistor $V_3$ is digitized through an amplifier and comparator circuit, where the binary output is sent to the MCU that acts as a digital feedback controller. The PLL systematically receives the prior output from an sMTJ and feeds it back into the device after processing through a proportional-integral-derivative (PID) controller (see Fig. 3(b)).

In Fig. 3(c) the blue curve shows the AP state probability from an sMTJ device with PLL, where $I_{SOT}$ (red curve) is used as the controlling signal to compensate any fluctuation in the environmental magnetic field over time. We see that shortly after the feedback is turned on, $p_{AP}$ stabilizes at 50 %. We even intentionally change the applied magnetic field from 0 to 3 Oe in the middle of the measurement to test the robustness of PLL, where $p_{AP}$ is observed to track the 50 % target very well during this transition. As a control experiment, we also measure the probability without PLL, and $p_{AP}$ is seen to be very susceptible to environmental fluctuations with a significant steady-state deviation from 50% under both applied fields, as shown in Fig. 3(d). In the open loop configuration, the sigmoid curves of the switching probability as a function of $I_{SOT}$ for $H = 0$ and 3 Oe are provided in the insets of Fig. 3(d), where the required currents for 50% probability align well with the $I_{SOT}$ signal determined by the PLL in Fig. 3(c). The comparison between Fig. 3(c) and 3(d) shows that the PLL effectively stabilizes the sMTJ, enabling sMTJs to generate random bits reliably even in the field-free regime.

Besides a stable 50% probability, practical RNGs must also control the fluctuation frequency (the inverse of characteristic dwell time), which is otherwise influenced by changes in ambient temperature and device variations from inconsistencies in patterning. Figure 3(e) shows the autocorrelation function of the random bitstream from the sMTJ. For statistically meaningful randomness, the random bits must be sampled with long enough time intervals that yield an autocorrelation sufficiently close to zero. Spatial or temporal variations in the fluctuation frequency will present substantial difficulties in the design of a robust system. Here, we introduce a frequency locked loop (FLL) to utilize $I_{MTJ}$ to control the fluctuation frequency (see Fig. 3(b)).

Figure 3(f) showcases a device with a stabilized fluctuation frequency at 250Hz after FLL is turned on, as measured in a room without precise control of the ambient temperature. The reference $I_{MTJ}$ versus $1/\tau$ curve measured in this open loop configuration (inset of Fig. 3(f)) aligns well with the current employed in the FLL results.

**True Randomness under NIST Tests**

We next demonstrate a field-free, single-sMTJ that passes the NIST SP 800-22 Statistical Test (NIST tests), a comprehensive test suite consisting of 15 different tests that assess the quality of randomness in the generated random bitstream [34]. While the PLL already stabilizes the AP state probability within a small window, the TRNG test has even more stringent criteria requiring $p_{AP}$ to be exceedingly close to 50 %. To further improve the accuracy on $p_{AP}$, an edge digitization (ED) method is introduced to process the output signal from sMTJ, as illustrated in Fig. 3(b). In contrast to conventional level digitization (LD), ED triggers a transition from '0' to '1' or from '1' to '0' on every detected rising edge. Conceptually, the sMTJ must undergo successive P→AP and AP→P transitions to switch the random bit under the ED scheme. Therefore, in ED the impact of residual asymmetry between the two energy wells is suppressed and the resultant telegraph noise is whitened. For circuit implementation, ED can be easily realized with simple digital logic gates, which do not significantly increase the system's footprint or energy consumption.

As shown in Table 1, the random bits generated by sMTJs with PLL and ED under a field-free regime pass all the tests and exhibit a clear distinction from other tests where either PLL and/or ED were not applied. As a reference, a random bitstream generated by a 32-bit linear feedback shift register (LFSR) implemented in a field-programmable gate array (FPGA) is also subjected to NIST tests. Due to the inherent mechanisms of LFSRs, they are unable to pass certain NIST tests, thereby exhibiting a degree of predictability [35]. Note that in this experiment, the combination of PLL and ED alone suffices to ensure the generation of high-quality random numbers, while the FLL primarily influences the optimal sampling frequency and is not employed. Our sMTJs generally have relatively large dwell time due to the large magnetic volume in the free layer, and a sample frequency of 250 Hz has been used to minimize the autocorrelation. We project that with lower barrier sMTJs fluctuating at the nanosecond time scale, as demonstrated in recent experiments [30,36], PLL and ED methods could further increase the sampling frequency without raising autocorrelation.

**Generative Diffusion Models with sMTJs**

Stochastic inputs are heavily utilized in contemporary machine learning techniques, *e.g.*, to achieve global optimization in the presence of local minima, or to create new data based on patterns learned from existing ones. In particular, generative AI models, including vision models (e.g., diffusion models [37]), language models (e.g., GPT-4 [38]), and multimodal models (e.g., DALL-E [39]), require random numbers as essential inputs for sampling from probability distributions to generate diverse, realistic, and novel outputs, and to capture the inherent variability and uncertainty in the data. Notably, it is crucial for random numbers to follow specific well-defined probability distributions to generate highly realistic outputs from such models.

We run a generative AI model using the random bitstream from sMTJs under field-free operation. Here, we demonstrate with two popular generative vision models: the denoising diffusion implicit model (DDIM) [37] and the deep convolutional generative adversarial network (DCGAN, Extended Data Fig. 3) [40]. DDIM, an accelerated version of the conventional diffusion model [41], starts from an image composed of pure noise and denoises it at each step using a non-Markovian process. After multiple steps, the noise within the image is finally eliminated, resulting in a meaningful picture. To generate the seeding noise in the initial image, we produce 8-bit precision random numbers based on the random bitstream from sMTJs (see Fig. 4(a)), in contrast to conventional realizations using software (Mersenne-Twister algorithm) or LFSRs. The 8-bit random numbers produced from sMTJs follow a uniform distribution since the original random bitstream maintains a 50% probability with low correlation between adjacent bits. Subsequently, a Box-Muller Transform [42] is used to convert these uniformly distributed random numbers to a standard normal distribution, as required in most AI algorithms including the initial inputs for the noisy image (see Fig. 4(b)). Here, we used the training dataset of CelebA-HQ resized by 256×256 [43]. While generative AI models usually do not have the highest requirements for all aspects of randomness as in the NIST test, random numbers reliably following a predefined target distribution with low autocorrelation remain crucial for ensuring the generated data's validity and meaningfulness. As demonstrated in Fig. 4(c), high quality image generation is only achieved with random bitstreams produced under PLL and ED, illustrating the effectiveness of our sMTJ TRNGs.

## Conclusions

We demonstrate the potential of three-terminal superparamagnetic tunnel junctions (sMTJs) as reliable and potentially scalable sources of true randomness under a field-free regime. By capitalizing on dual-current controllability and feedback system including probability locked loop (PLL) and frequency locked loop (FLL), we delineate effective methods to augment the stability and reliability of sMTJ-based systems under fluctuating conditions, addressing the variability and scalability issues intrinsic to low energy barrier sMTJs. The generation of high-quality random bitstreams meeting all NIST test standards, and their successful deployment in generative artificial intelligence models underscore the broad applicability of this technology in advanced computing. Overall, this work charts a course for the robust and scalable application of sMTJs in cryptography, machine learning at edge devices and beyond, addressing critical challenges such as device variability, environmental sensitivity and magnetic field requirement.

## Acknowledgements

The authors thank Zhiping He for assistance in wire bonding devices. D.K. thanks Seou Choi and Chanwoo Park for helpful discussion on generative diffusion models. D.K. thanks Adam Kim and Dongchel Shin for helpful discussion on control theory. The work is partly supported by Semiconductor Research Corporation (SRC) and DARPA. B.C.M. was supported by National Science Foundation under award DMR-2104912.

## Author contributions

D.K., Q.W., B.C.M., L.L., and M.A.B. conceived and designed the project. L.L. and M.A.B. supervised the project. D.K. and B.C.M. performed materials characterization, fabricated and optimized the sMTJ devices. D.K., Q.W. and B.C.M. wrote the codes for measurement and analysis. D.K. and C.-T.C. devised the concept of edge digitization. D.K. and Q.W. designed the peripheral circuits. D.K., Q.W. and B.C.M. performed electrical measurement and analysis. D.K. completed data processing and wrote the manuscript. L.L. and M.A.B. edited the manuscript. All the authors participated in the discussion and analysis of the manuscript.

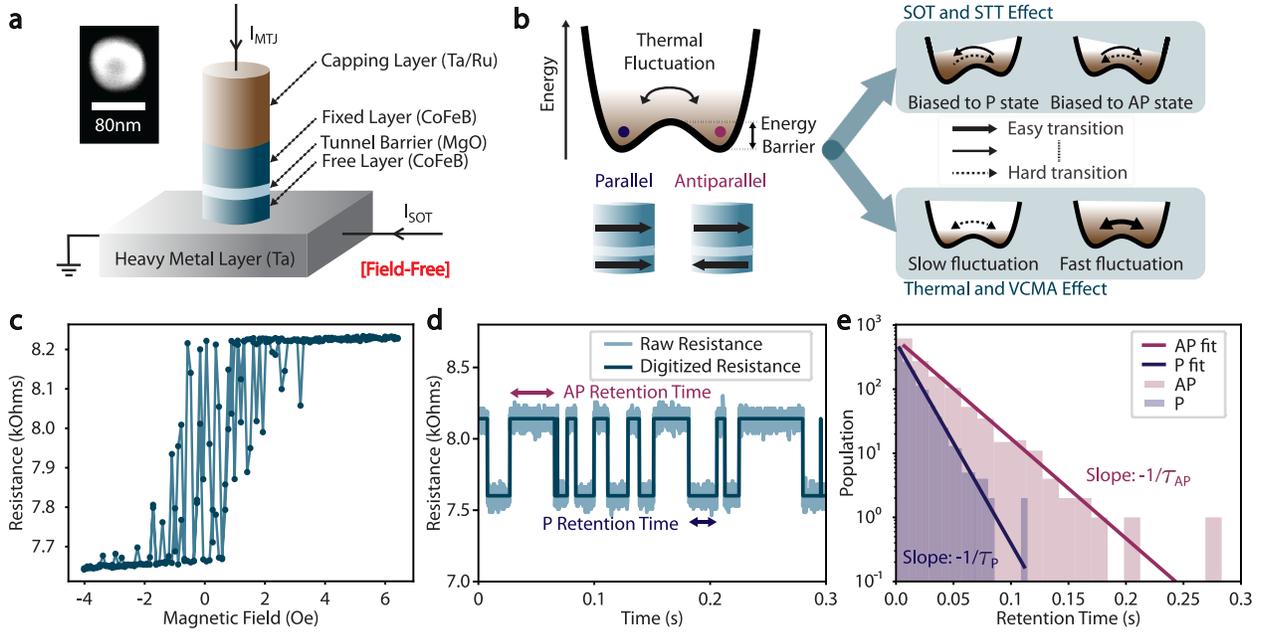

**Fig. 1.** (a) Schematic diagram of a three-terminal sMTJ noting the directions of the biasing currents. The inset shows the scanning electron micrograph image of a nanopillar with elliptical size of 77 nm × 73 nm. (b) Double-well energy landscape. The energy landscape can be controlled by biasing currents via spin-orbit-torque (SOT), spin-transfer-torque (STT), thermal, and voltage-controlled magnetic anisotropy (VCMA) effects. (c) Minor loop of the sMTJ demonstrating stochastic behavior of the device. (d) Time trace of thermally activated bipolar fluctuation between parallel (P) and antiparallel (AP) states. (e) Exponential distribution of P and AP retention times. Dwell times $\tau_{P,AP}$ are determined as their time constants.

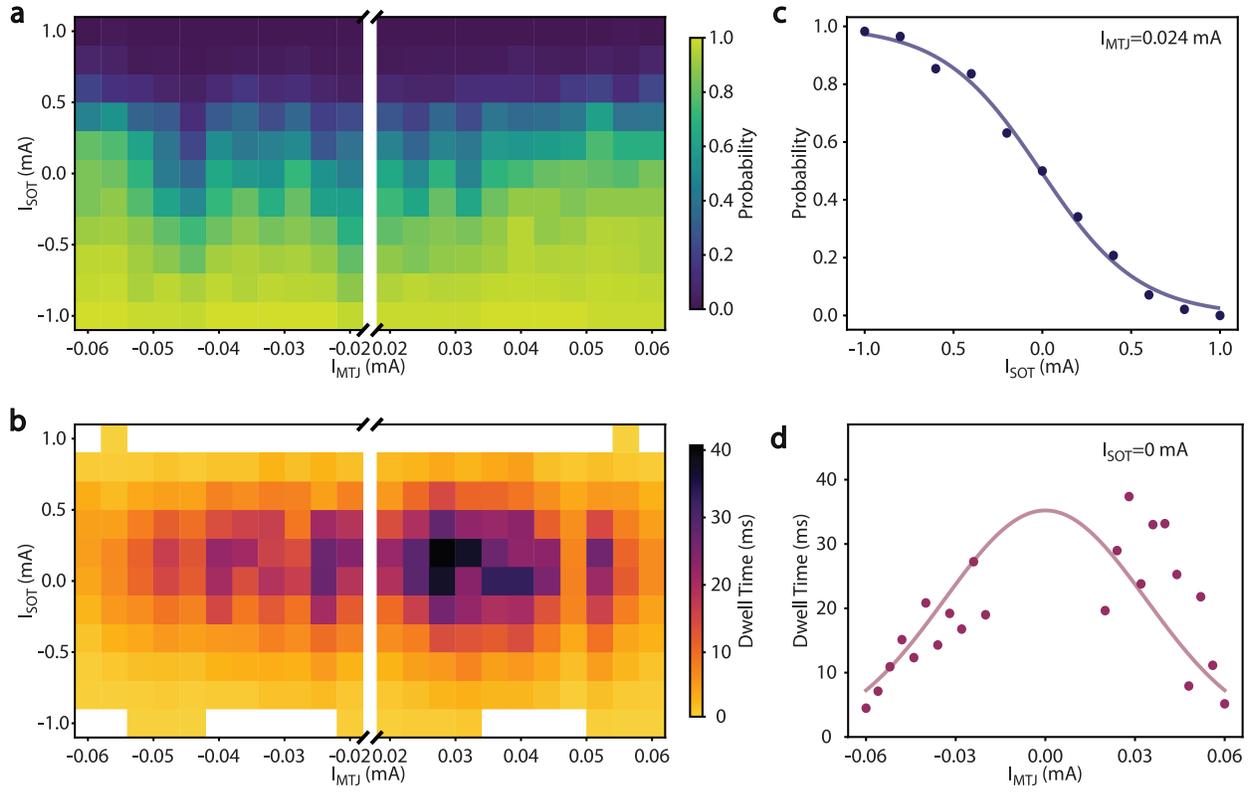

**Fig. 2.** Dual-current controllability of (a) probability of the AP state and (b) characteristic dwell time under 1 Oe. (c) and (d) represents example cuts along $I_{SOT}$ and $I_{MTJ}$ directions in (a) and (b), separately. In (c) with a fixed $I_{MTJ}$, the probability exhibits a sigmoidal dependence on $I_{SOT}$. In (d) with a fixed $I_{SOT}$, the characteristic dwell time approximately follows a Gaussian dependence on $I_{MTJ}$.

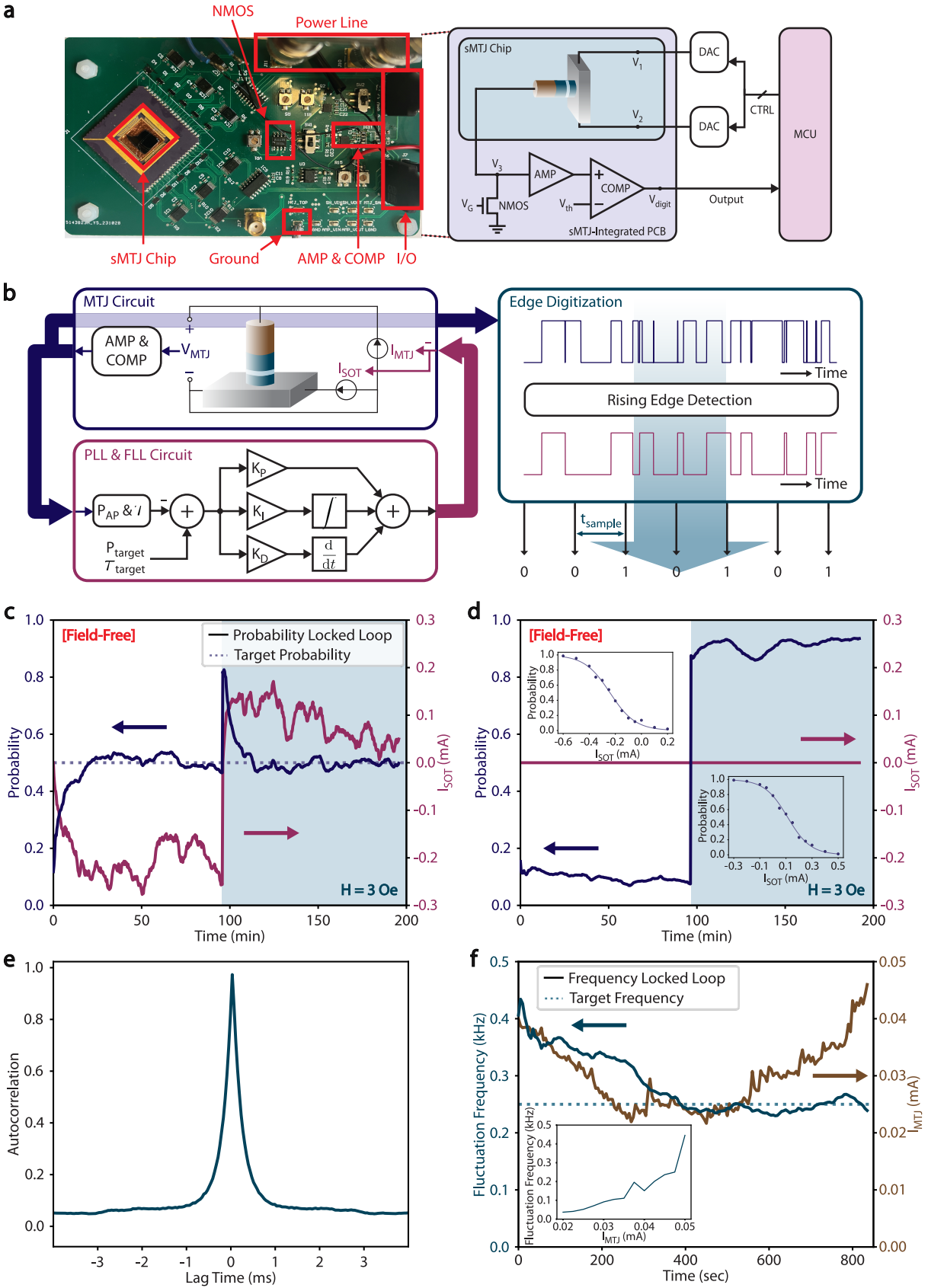

**Fig. 3.** (a) Interface circuit between a sMTJ chip and a microcontroller unit (MCU) implemented on a printed circuit board (PCB). (b) Schematic diagram of an sMTJ-based system with a probability locked loop (PLL), a frequency locked loop (FLL), and edge digitization (ED) enabled. A proportional-integral-derivative (PID) controlling circuit is used as part of PLL and FLL circuits. The ED is realized by rising edge detection in the MCU. (c) and (d) Changes in probability (c) with and (d) without PLL. In the middle of both the two tests, the magnetic field is abruptly changed from 0 (field-free) to 3 Oe. The target probability for PLL is 50%. The insets in (d) show open loop $p_{AP}$ vs $I_{SOT}$ for applied field of 0 and 3 Oe, separately. (e) Autocorrelation with respect to lag time. Random numbers should be sampled periodically with a sampling frequency that achieves autocorrelation sufficiently close to zero. (f) Changes in fluctuation frequency in the sMTJ with FLL. The target frequency for FLL is 250 Hz, with the inset in (f) showing the fluctuation frequency vs $I_{MTJ}$ relation under the ambient temperature of around 15°C.

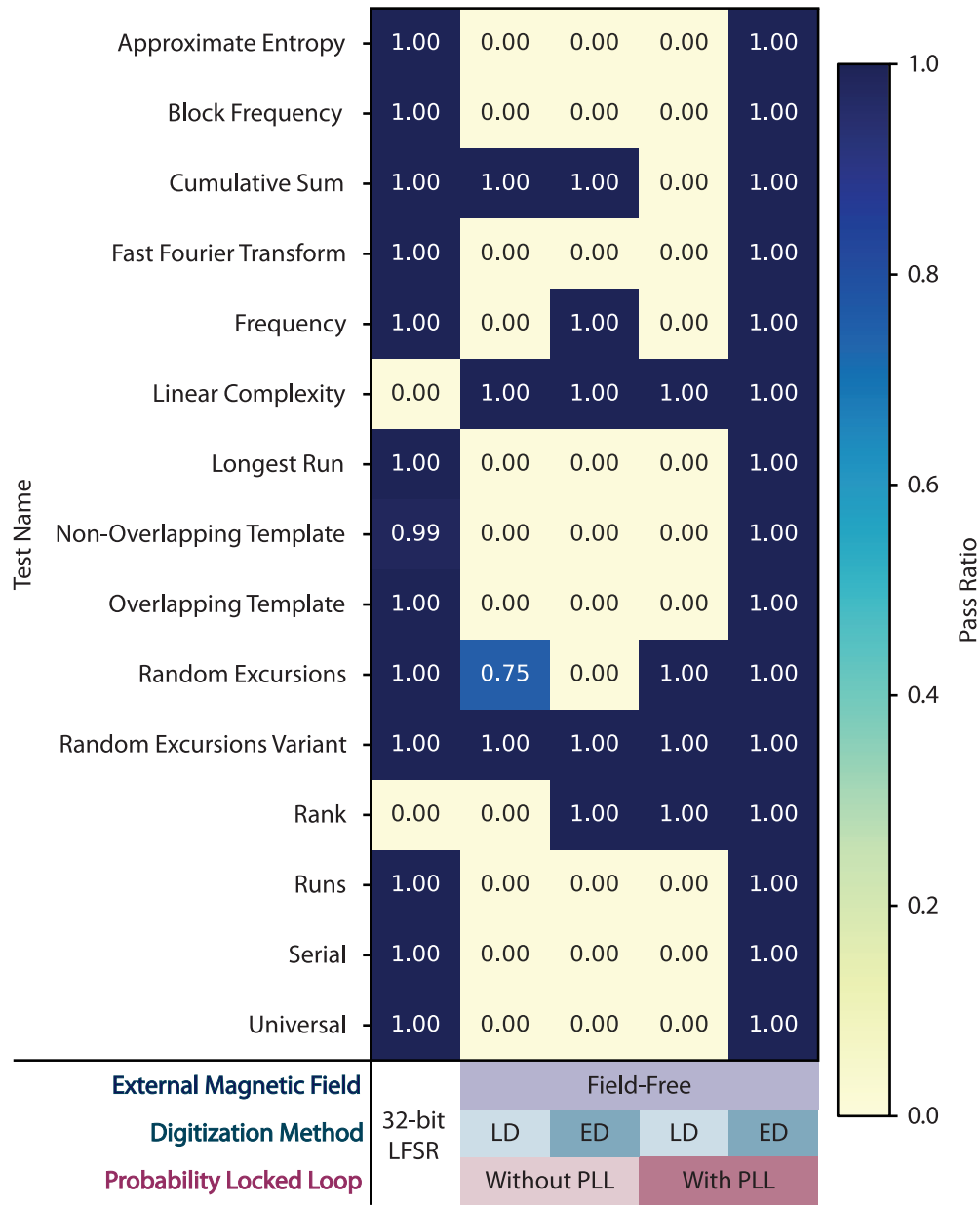

| Test Name | 32-bit LFSR | LD | ED | LD | ED |
|---|---|---|---|---|---|
| Approximate Entropy | 1.00 | 0.00 | 0.00 | 0.00 | 1.00 |
| Block Frequency | 1.00 | 0.00 | 0.00 | 0.00 | 1.00 |
| Cumulative Sum | 1.00 | 1.00 | 1.00 | 0.00 | 1.00 |
| Fast Fourier Transform | 1.00 | 0.00 | 0.00 | 0.00 | 1.00 |
| Frequency | 1.00 | 0.00 | 1.00 | 0.00 | 1.00 |
| Linear Complexity | 0.00 | 1.00 | 1.00 | 1.00 | 1.00 |
| Longest Run | 1.00 | 0.00 | 0.00 | 0.00 | 1.00 |
| Non-Overlapping Template | 0.99 | 0.00 | 0.00 | 0.00 | 1.00 |
| Overlapping Template | 1.00 | 0.00 | 0.00 | 0.00 | 1.00 |
| Random Excursions | 1.00 | 0.75 | 0.00 | 1.00 | 1.00 |
| Random Excursions Variant | 1.00 | 1.00 | 1.00 | 1.00 | 1.00 |
| Rank | 0.00 | 0.00 | 1.00 | 1.00 | 1.00 |
| Runs | 1.00 | 0.00 | 0.00 | 0.00 | 1.00 |
| Serial | 1.00 | 0.00 | 0.00 | 0.00 | 1.00 |
| Universal | 1.00 | 0.00 | 0.00 | 0.00 | 1.00 |
| **External Magnetic Field** | | Field-Free | | | |
| **Digitization Method** | | LD | ED | LD | ED |
| **Probability Locked Loop** | | Without PLL | | With PLL | |

**Table 1.** NIST Statistical Test Suite results with and without PLL and ED under a field-free regime. The vertical axis lists various tests in the suite. A 32-bit LFSR implemented in a FPGA is tested as a control experiment. Each number in the cells indicates the pass ratio. When the PLL and ED are both applied, the random bits generated from sMTJ possess true randomness in the given sampling frequency (250 Hz), while each method alone is not sufficient for true randomness under the same sampling frequency.

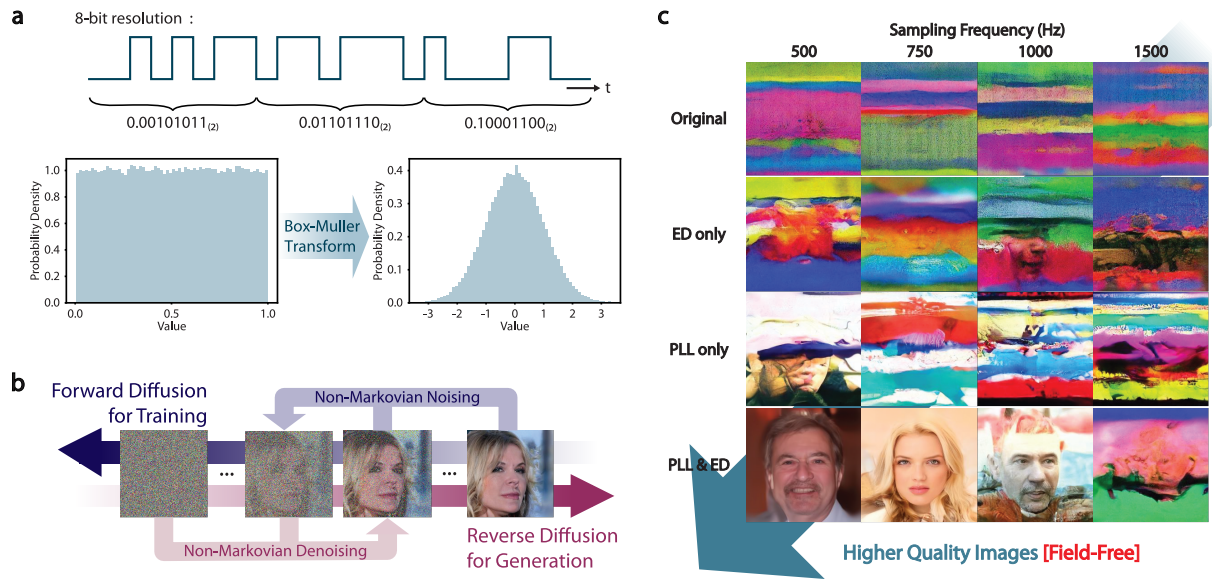

**Fig. 4.** (a) Translation of a random bitstream into a standard normal distribution. Initially, 8 bits of random bits are grouped to form a binary fraction. These numbers are then processed through the Box-Muller Transform. If the input distribution is uniform, the output will be a standard normal distribution. (b) Schematic diagram of denoising diffusion implicit model (DDIM). Starting with a normal-distributed noisy image, DDIM utilizes a non-Markovian process for image denoising and generation. (c) Images generated by DDIM using field-free random bit generation from sMTJ. PLL and ED methods with lower sampling frequencies yield better images, due to the higher quality of randomness.

# Methods

**sMTJ Fabrication**

A material stack consisting of (from bottom to top) Ta (14 nm)/CoFeB (1.3 nm)/MgO (1 nm)/CoFeB (2.6 nm)/Ta (5 nm)/Ru (5 nm) is deposited with the magnetron sputtering system on a silicon wafer equipped with the 1-μm-thick oxide layer. It is then subjected to a thermal annealing at 300°C for 5 minutes. Electron-beam (e-beam) lithography is employed to define tantalum current channels (1 μm wide) and tunnel junctions (approximately 75 nm in diameter). Each junction is shaped as an ellipse with a low aspect ratio, so that the magnetic free layer possesses two states separated by a low energy barrier. The volume is also minimized for the given thickness of the CoFeB free layer to further lower down the thermal stability. An ion miller equipped with an end-point detector is utilized to etch the material stack down to the exact depth. The pillar is then electrically isolated with silicon oxide deposition using e-beam evaporator. Finally, contact pads are formed with a stack of (from bottom to top) Ti (10 nm)/Cu (55 nm)/Au (10 nm), also by e-beam lithography for patterning and e-beam evaporator for deposition.

**Discrete Time Proportional-Integral-Derivative (PID) Control**

PID feedback control is used to stabilize the probability of the AP state and the fluctuation frequency of the sMTJs. For a probability locked loop (PLL), the probability of the AP state $p_{AP}$ is fixed to 50%, which is the target probability $p_{target}$. At time $n$, assume that the $I_{SOT}[n]$ is applied and as a result the probability $p_{AP}[n]$ is obtained. Then

$$E[n] = p_{target} - p_{AP}[n]$$

$$I_{SOT}[n+1] = I_{SOT}[n] - K_p E[n] - K_i \sum_{i=1}^{n} E[i] - K_d (E[n] - E[n-1])$$

when $n \geq 1$. Here, $E[n]$ is the error and $I_{SOT}$ is defined in the main text. One set of the working PID parameters are experimentally determined to be $K_p = 0.05$ (mA), $K_i = 0.001$ (mA), and $K_d = 0.02$ (mA). This feedback loop effectively pins the probability of sMTJ to the target probability 50%.

In the frequency locked loop (FLL) the characteristic dwell time is controlled with a target dwell time of $\tau_{target} = 4$ (ms). At time $n$, assume that the control current $I_{MTJ}[n]$, defined in the main text, is applied and as a result the dwell time $\tau[n]$ is obtained. Then,

$$E[n] = \tau_{target} - \tau[n]$$

$$I_{MTJ}[n+1] = I_{MTJ}[n] - K_p E[n] - K_i \sum_{i=1}^{n} E[i] - K_d (E[n] - E[n-1])$$

when $n \geq 1$. One set of the working PID gains are determined to be $K_p = 0.2$ (mA/s), $K_i = 0.008$ (mA/s), and $K_d = 0.6$ (mA/s).

The operation of the sMTJs within the printed circuit board (PCB) can be directly controlled by the voltages $V_1, V_2$ and $V_G$ (see Fig. 3(a)). As the scale of $I_{MTJ}$ is much smaller than that of $I_{SOT}$, $I_{SOT} \approx (V_1 - V_2)/R_{ch}$, where $R_{ch}$ is the tantalum channel resistance. Also, by modulating the gate voltage of NMOS ($V_G$) with fixed $V_{DD} = (V_1 + V_2)/2$, the $I_{MTJ}$ can be effectively controlled using the transfer curve of the NMOS.

**Linear Feedback Shift Register (LFSR) Implementation on Field-Programmable Gate Array (FPGA)**

A 32-bit LFSR is implemented on a FPGA to compare its random number quality with the ones collected from sMTJs. The feedback polynomial used is to maximize the non-repeatable pattern length [44]. The LFSR is memory-mapped to a softcore RISC-V central processing unit (CPU) on FPGA, which communicates with a personal computer (PC) using a serial port. When collecting the stochastic bitstream, a 32-bit random seed generated by PC software is used to initialize the LFSR, after which 1.6 million consecutive stochastic bits from LFSR are recorded by the PC.

**Box-Muller Transform**

Let random variables $U_1$, $U_2$ follow a uniform distribution ranging from 0 to 1 $\mathcal{U}(0,1)$, and let random variables $Z_0$ and $Z_1$ be

$$Z_0 = \sqrt{-2\ln U_1}\cos(2\pi U_2), Z_1 = \sqrt{-2\ln U_1}\sin(2\pi U_2),$$

then according to the Box-Muller Transform, the random variables $Z_0$ and $Z_1$ are independent and follow a standard normal distribution $\mathcal{N}(0,\ 1^2)$ [42].

The Box-Muller Transform and its variants are widely used algorithms for generating normal distribution in conventional computers [45]. This work followed the same procedure that modern generative artificial intelligence models use, by first generating a random bitstream with a 50% probability, pairing them up, and then applying the Box-Muller Transform to produce random numbers following the standard normal distribution. In this work, pairs of 8-bit random bits are also fed to the Box-Muller Transform as $U_1$, $U_2$. 8-bit random bits follow a uniform distribution only when the random bitstream has a 50% probability and each bit is not correlated with adjacent bits. Hence, it is crucial to maintain a uniform distribution for random number generators to run algorithms that utilize normal distribution, including generative models.

**Denoising Diffusion Models**

Denoising diffusion models (DDPMs [41] and DDIMs [37]) generate data through an iterative process that involves multiple steps. They consist of two processes: forward diffusion process for training by noising and reverse diffusion process for generation by denoising [37,41]. In this work, a pretrained network of DDPM with CelebA-HQ 256 dataset is used [41], but the image generation is done with DDIM [37]. To accelerate generation, the reverse process is engineered to skip some intermediate steps, compressing the total number of steps to 50. Random numbers generated from sMTJs under field-free condition are fed to the model as initial Gaussian noisy images, after the conversion to standard normal distribution using Box-Muller Transform.

**Generative Adversarial Networks (GANs)**

In Extended Fig. 3, deep convolutional generative adversarial networks (DCGANs) [40] are utilized. In DCGANs, both generator and discriminator consist of deep convolutional layers instead of fully connected layers [40]. This work uses a standard normal distribution as the latent vector distribution. Therefore, Gaussian noise should be input to the generator to generate meaningful images. Pretrained DCGANs [46] with MNIST [47] and CIFAR-10 [48] datasets are used. The size of latent vector is 100.

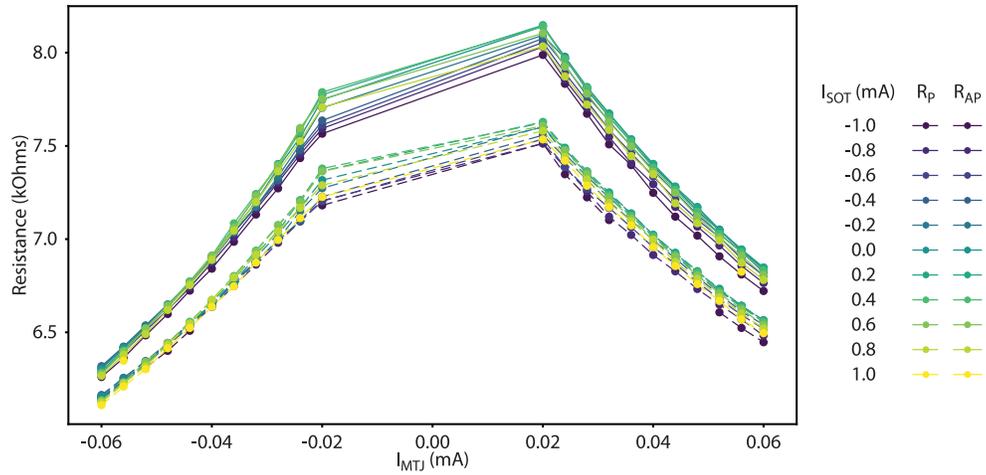

**Extended Data Fig. 1.** Resistance of parallel ($R_P$) and antiparallel ($R_{AP}$) states across different current biasing. Tunneling magnetoresistance tends to be higher under lower voltage difference across the sMTJ nanopillar.

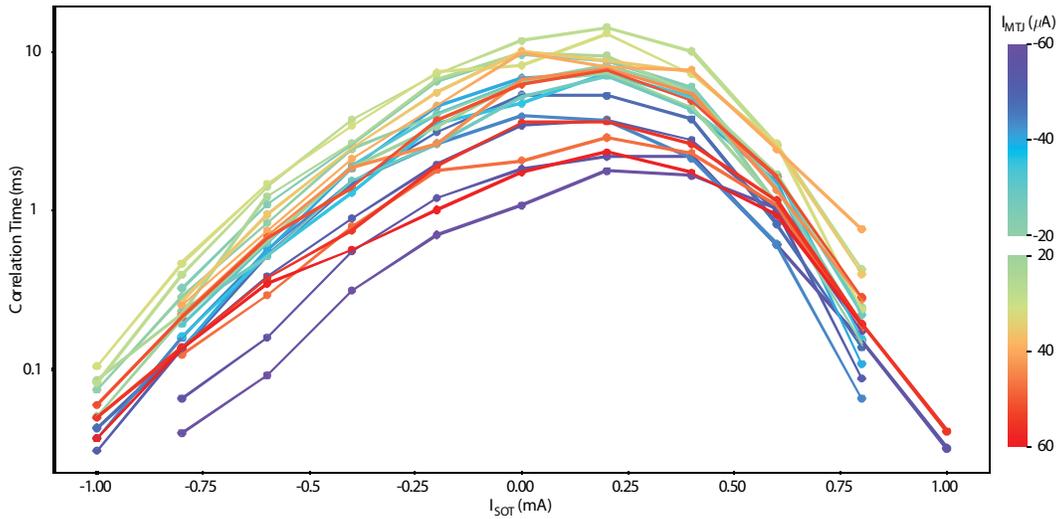

**Extended Data Fig. 2.** Correlation time versus $I_{SOT}$ relationship for each $I_{MTJ}$ biasing condition. Correlation time is defined as the lag time when the autocorrelation crosses 0.5. For extreme $I_{SOT}$ biases the sMTJ is deterministic during the measurement window, therefore the autocorrelation is less meaningful.

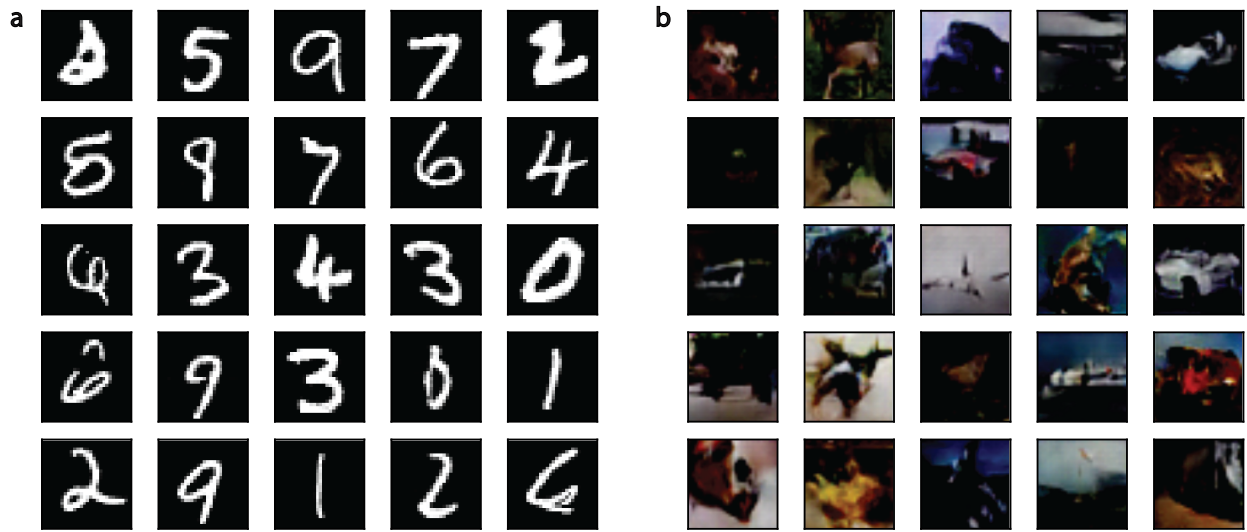

**Extended Data Fig. 3.** Generated images from a deep convolutional generative adversarial network (DCGAN) trained with (a) MNIST and (b) CIFAR-10 using sMTJs under field-free operation with probability locked loop (PLL) and edge digitization (ED) methods.